\documentclass[12pt]{article}
\usepackage{amsmath}
\def\a{\alpha}

\def\s{\sigma}

\def\be{\begin{equation}}
\def\ee{\end{equation}}
\def\arr{\begin{array}{rll}}
\def\ea{\end{array}}
\def\bea{\begin{eqnarray}}
\def\eea{\end{eqnarray}}
\def\N2{$N{=}2$}

\def\sfrac#1#2{{\textstyle\frac{#1}{#2}}}
\def\>{\rangle}
\def\<{\langle}
\def\+{\dagger}
\def\={\ =\ }

\def\cJ{\mathcal{J}}
\def\bal{\begin{aligned}}
\def\eal{\end{aligned}}
\begin{document}
\begin{titlepage}
\setcounter{page}{0}
\begin{center}
{\Large\bf  Spinning particles on $\mathcal{S}^2$ in accord }\\
\vskip 0.4cm
{\Large\bf with the Bianchi classification }\\
\vskip 1.5cm
\textrm{\Large Anton Galajinsky \ }
\vskip 0.7cm
{\it
Tomsk Polytechnic University, 634050 Tomsk, Lenin Ave. 30, Russia} \\
\vskip 0.2cm
{e-mail: galajin@tpu.ru}
\vskip 0.5cm
\end{center}

\begin{abstract} \noindent
Motivated by recent studies of superconformal mechanics extended by spin degrees of freedom, we
construct minimally superintegrable models of spinning particles on $\mathcal{S}^2$, the spin degrees of freedom of which are represented by a $3$--vector obeying the structure relations of a $3d$ real Lie algebra. Generalisations involving an external field of the Dirac monopole, or the motion on the group manifold of $SU(2)$, or a scalar potential giving rise to two quadratic constants of the motion are discussed.
A procedure how to build similar extensions, which rely upon $d=4,5,6$ real Lie algebras, is elucidated.
\end{abstract}

\vspace{0.5cm}

\end{titlepage}
\renewcommand{\thefootnote}{\arabic{footnote}}
\setcounter{footnote}0

\noindent
{\bf 1. Introduction}\\

\noindent
Over the last few decades, models of superconformal mechanics attracted a considerable amount of attention (for a review see \cite{FIL}). For one thing, they proved useful for describing a super $0$--brane propagating on a near horizon extreme black hole background as well as for a microscopic description of the latter. For another thing,
they are relevant for understanding the $AdS_2/CFT_1$–-correspondence.

More recently, the focus of research shifted to the study of superconformal mechanics extended by spin degrees of freedom \cite{FIL1}--\cite{F1}.
Such variables typically arise when gauging $U(n)$ isometry of the matrix superfield systems \cite{FIL1} or supersymmetrizing the Euler--type extension of the Calogero model \cite{KLS}.
Because the new variables entail a richer structure of admissible couplings, one can bypass some long--standing problems. To mention a few, the spin--extended superconformal models can be formulated for an arbitrary number of particles and they can accommodate an arbitrary even number of supersymmetries in a way compatible with non--trivial interactions \cite{KLS,KLS1}.

In a recent work \cite{GL}, an alternative approach of introducing spin degrees of freedom was advocated, which promoted a dynamical realization of
$su(2)$ associated with the model of a relativistic spinning particle propagating on a spherically symmetric curved background to that of the $D(2,1;a)$ superconformal mechanics.\footnote{The exceptional supergroup $D(2,1;a)$ describes the most general $\mathcal{N}=4$ supersymmetric extension of the conformal group in one dimension $SO(2,1)$ (see e.g. the discussion in \cite{FIL}). The structure relations of the corresponding Lie superalgebra involve a real parameter $a$.} The spin variables were represented by a symmetric Euler top.

Within the Hamiltonian formalism, the Euler top is usually described by the angular velocity vector $J_i$, $i=1,2,3$, obeying the $su(2)$ structure relations $\{J_i,J_j \}=\epsilon_{ijk} J_k$ under the Poisson bracket. Classification of $3d$ real Lie algebras was accomplished in \cite{B}, where $su(2)$ was identified with the type--IX algebra. One may wonder whether the construction in \cite{GL} can be generalised to cover other instances from the Bianchi classification.

The aim of this paper is to construct a minimally superintegrable spinning particle on $\mathcal{S}^2$, the spin degrees of freedom of which are represented by a $3$--vector obeying the structure relations of a $3d$ real Lie algebra. An extension to the $D(2,1;a)$ superconformal mechanics is then straightforward \cite{G1}, which would result in a supermultiplet of the type $(3,4,1)$ accompanied by the spin degrees of freedom.\footnote{In modern literature, it is customary to associate the non--relativistic spin with the $SU(2)$ group. Other options are usually referred to as "internal degrees of freedom". In this work, we loosely use the term "spin variables" irrespectively of a $3d$ real Lie algebra at hand.}

The work is organised as follows. In Sec. 2, we briefly remind how a dynamical realization of $su(2)$ on some Poisson manifold can be extended to accommodate the $D(2,1;a)$ superconformal symmetry. 
In Sec. 3, minimally superintegrable spinning particles on $\mathcal{S}^2$ are constructed and ranked in accord with the Bianchi classification \cite{B}. The construction includes a few steps. First, one
chooses a $3d$ real Lie algebra with generators $J_i$, $i=1,2,3$, and structure constants $c_{ij}^k$ and identifies $J_i$ with the spin degrees of freedom obeying the (degenerate) Poisson bracket
$\{J_i,J_j \}=c_{ij}^k J_k$. Then one decomposes the spin vector $\vec{J}$ on the orthonormal frame attached to a point on $\mathcal{S}^2$ embedded in $\mathcal{R}^3$ and extends the conventional angular momentum vector $\vec{L}$ of a free particle on $\mathcal{S}^2$ to include the spin part $\vec{\cJ}=\vec{L}+\vec{J}$. Afterwards, the Poisson brackets among the momenta canonically conjugate to the angular variables and the spin degrees of freedom are fixed from the requirement that $\vec{\cJ}$ obeys the $su(2)$ structure relations, the condition that the Casimir element of a $3d$ real Lie algebra be an integral of motion of a dynamical system governed by the Hamiltonian $H=\frac 12 \vec{\cJ}^2$, and the fulfilment of the Jacobi identities. In Sec. 4, a qualitative dynamical behaviour of the systems is analysed and another perspective on
the material in Sec. 3 is offered. A procedure how to build similar minimally superintegrable extensions, which rely upon $d=4,5,6$ real Lie algebras, is elucidated.
Generalisations of the models in Sec. 3, which are compatible with the minimal superintegrability, are given in Sec. 5. In the concluding Sec. 6, we summarise our results. Poisson brackets among the momenta canonically conjugate to the angular variables and the spin degrees of freedom are gathered in Appendix.

\vspace{0.5cm}

\noindent
{\bf 2. Extending a dynamical realization of $su(2)$ to that of $D(2,1;a)$}\\

\noindent
Following Ref. \cite{G1}, let us briefly remind how a dynamical realization of $su(2)$ on a Poisson manifold can be extended to accommodate the $D(2,1;a)$ superconformal symmetry.

Let a Poisson manifold be parametrized by real variables $\Gamma_\alpha$, $\alpha=1,\dots,d$, obeying the bracket $\{\Gamma_\alpha,\Gamma_\beta \}=\Omega_{\alpha\beta}(\Gamma)$. It is assumed that $\Omega_{\alpha\beta}(\Gamma)=-\Omega_{\beta\alpha}(\Gamma)$ and the Jacobi identities are satisfied. In general, $\Omega_{\alpha\beta}(\Gamma)$ is allowed to be a degenerate matrix.

Consider three functions $J_i(\Gamma)$, $i=1,2,3$, which obey the $su(2)$ structure relations
\be\label{su(2)}
\{J_i,J_i \}=\epsilon_{ijk} J_k,
\ee
where $\epsilon_{ijk}$ is the Levi--Civita symbol with $\epsilon_{123}=1$.
The Lie superalgebra associated with the exceptional superconformal group $D(2,1;a)$ involves two $su(2)$ subalgebras, the first of which is identified with $R$--symmetry, while the second transforms fermions only. Below, $J_i$ will enter the $R$--symmetry generator.

In order to incorporate (\ref{su(2)}) into a dynamical realization of $D(2,1;a)$, it suffices to extend $\Gamma_\alpha$ by bosonic canonical variables $(x,p)$ and a pair of complex conjugate $SU(2)$--spinors $\psi_\alpha$, $\bar\psi^\a={(\psi_\alpha)}^{*}$, $\alpha=1,2$, which satisfy the brackets
\be\label{br}
\{x,p\}=1, \qquad \{ \psi_\alpha, \bar\psi^\beta \}=-{\rm i} {\delta_\alpha}^\beta.
\ee
On such an extended Poisson supermanifold one then considers the set of functions
\begin{align}\label{rep}
&
H=\frac{p^2}{2}+\frac{2 a^2 }{x^2} J_i J_i+\frac{2 a}{x^2} (\bar\psi \s_i \psi) J_i -\frac{(1+2a)}{4x^2} \psi^2 \bar\psi^2, && D=tH-\frac 12 x p,
\nonumber\\[2pt]
&
K=t^2 H-t x p +\frac 12 x^2, && \mathcal{L}_i=J_i+\frac 12 (\bar\psi \s_i \psi),
\end{align}
\begin{align}
&
Q_\a=p \psi_\a-\frac{2{\rm i} a}{x} {(\s_i \psi)}_\a J_i -\frac{{\rm i}(1+2a)}{2x} \bar\psi_\a \psi^2\ , && S_\a=x \psi_\a -t Q_\a,
\nonumber\\[2pt]
&
\bar Q^\a =p \bar\psi^\a+\frac{2{\rm i}a}{x} {(\bar\psi \s_i)}^\a J_i -\frac{{\rm i}(1+2a)}{2x} \psi^\a \bar\psi^2, &&
\bar S^\a=x \bar\psi^\a -t \bar Q^\a,
\nonumber\\[2pt]
&
I_{-}=\frac{{\rm i}}{2} \psi^2, \qquad \qquad \qquad \qquad I_{+}=-\frac{{\rm i}}{2} {\bar\psi}^2, &&
I_3=\frac 12 \bar\psi \psi,
\end{align}
where $a$ is an arbitrary real parameter,
$\psi^2=\psi^\a \psi_\a$, $\bar\psi^2=\bar\psi_\a \bar\psi^\a$, $\bar\psi\psi=\bar\psi^\a \psi_\a$, and ${{\left(\s_i\right)}_\alpha}^\beta$ are the Pauli matrices. These prove to reproduce the structure relations of the Lie superalgebra corresponding to $D(2,1;a)$ under the Poisson bracket chosen \cite{G1}. $H$ is the Hamiltonian of the resulting dynamical system. $D$ and $K$ are the generators of dilatations and special conformal transformations. $Q_\a$ and $S_\a$ are linked to supersymmetry transformations and superconformal boosts, while $\mathcal{L}_i$ and $I_{\pm}$, $I_3$ generate two $su(2)$ subalgebras.

The simplest realization of (\ref{su(2)}) is provided by the angular momentum vector of a free particle on $\mathcal{S}^2$, in which case (\ref{rep}) describes an on-shell $(3,4,1)$ supermultiplet. Vector fields dual to the conventional left--invariant one--forms on $SU(2)$ group manifold give rise to an off--shell $(4,4,0)$ supermultiplet. By properly adjusting $su(2)$ generators characterising a relativistic spinning particle propagating on a spherically symmetric curved background, one can achieve an extension of such supermultiplets by $SU(2)$--spin variables \cite{GL}.

In the next section, we construct a spinning particle on $\mathcal{S}^2$, the spin degrees of freedom of which are represented by a $3$--vector obeying the structure relations of a generic $3d$ real Lie algebra. Making use of the extended framework (\ref{rep}), one can automatically build the corresponding spinning extension of the $(3,4,1)$ supermultiplet.

\vspace{0.5cm}

\noindent
{\bf 3. Spinning particles on $\mathcal{S}^2$ in accord with the Bianchi classification}\\

\noindent
A group--theoretic description of a free particle on $\mathcal{S}^2$ identifies the geodesic Hamiltonian $H=\frac 12 g^{ij}p_i p_j$ with the Casimir element of $su(2)$ represented by the angular momentum vector $\vec{L}$
\be\label{H}
H=\frac 12 \vec{L}^2, \qquad
\vec{L}=
\begin{pmatrix} -p_\theta \sin{\phi}-p_\phi \cot{\theta} \cos{\phi}, \\ p_\theta \cos{\phi}-p_\phi \cot{\theta} \sin{\phi} \\ p_\phi \end{pmatrix},
\ee
where $(\theta,p_\theta)$ and $(\phi,p_\phi)$ are canonical pairs obeying the conventional Poisson brackets
\be\label{br0}
\{\theta,p_\theta\}=1, \qquad \{\phi,p_\phi\}=1.
\ee

In order to build a spinning extension of (\ref{H}), let us consider a $3d$ real Lie algebra with generators $J_i$, $i=1,2,3$, and structure constants $c_{ij}^k$ and identify $J_i$ with the spin degrees of freedom obeying the bracket
\be
\{J_i,J_j \}=c_{ij}^k J_k.
\ee
Classification of $3d$ real Lie algebras dates back to the work of Bianchi \cite{B}. The available options are displayed below in Table 1 (we follow a modern exposition in \cite{DNF}), which also contains the Casimir invariant for each case.
\newpage
\begin{center}
Table 1. The Bianchi classification of $3d$ real Lie algebras
\end{center}
\begin{eqnarray*}
\footnotesize
\begin{array}{|l|r|r|r|r|r|r|r|r|c|}
\hline
  & \{J_1,J_2 \}  & \{J_1,J_3 \}   & \{J_2,J_3 \}  & \mbox{Casimir element}~ \mathcal{I} \\
  \hline
~ \mbox{type I} & 0 & 0 & 0 &  J_1, J_2, J_3 \\
\hline
~ \mbox{type II} & 0 & 0 & J_1 &  J_1 \\
\hline
\mbox{ type III} & J_2-J_3  & -J_2+J_3 & 0 & J_2+J_3\\
\hline
\mbox{ type IV} & J_2+J_3 & J_3 & 0 & \frac{J_2}{J_3}-\ln{J_3} \\
\hline
\mbox{ type V} & J_2 & J_3 & 0 & \frac{J_2}{J_3}\\
\hline
\mbox{ type VI} & a J_2-J_3 & -J_2+a J_3 & 0 & J_3^2 {\left(1+\frac{J_2}{J_3} \right)}^{1+a} {\left(1-\frac{J_2}{J_3} \right)}^{1-a} \\
\hline
~ \mbox{type $$VI$_0$} & 0 & J_2 & J_1& J_1^2-J_2^2  \\
\hline
 \mbox{ type VII} & a J_2+J_3 & -J_2+a J_3 & 0 & (J_2^2 + J_3^2) e^{-2 a \arctan{\frac{J2}{J3}}}  \\
\hline
\mbox{ type $$VII$_0$} & 0 & -J_2& J_1 & J_1^2+J_2^2  \\
\hline
\mbox{ type VIII} & -J_3& -J_2 & J_1 & J_1^2+J_2^2-J_3^2  \\
\hline
 \mbox{ type IX} & J_3 & -J_2 & J_1 & J_1^2+J_2^2+J_3^2  \\
\hline
\end{array}
\end{eqnarray*}
In what follows we assume that $J_i$ commute with $(\theta,\phi)$
\be
\{\theta,J_i \}=0, \qquad \{\phi,J_i \}=0,
\ee
while the brackets $\{p_\theta,p_\phi \}$, $\{p_\theta,J_i \}$, $\{p_\phi,J_i\}$ will be fixed below (see (\ref{br3})). The abelian type--I case will be disregarded as it is of little physical interest. 

Note that it is customary nowadays to use such a formalism for the Hamiltonian description of rigid body dynamics. For example, focusing on $su(2)$ for which $c_{ij}^k=\epsilon_{ijk}$ and choosing the Hamiltonian
in the form $H=\frac 12 \left(g_1^2 J_1^2+g_2^2 J_2^2+g_3^2 J_3^2 \right)$, where $(g_1,g_2,g_3)$ are constants (moments of inertia), one can represent the Euler top equations as ${\dot J}_i=\{J_i,H\}$. Further examples of such a kind can be found in \cite{AP}.

As the next step, one considers a unit two--sphere embedded in $\mathcal{R}^3$, builds an orthonormal frame at each point\footnote{Given the parametric representation $x=\sin{\theta} \cos{\phi}$, $y=\sin{\theta} \sin{\phi}$, $z=\cos{\theta}$, one computes two vectors $(x'_\theta,y'_\theta,z'_\theta)$, $(x'_\phi,y'_\phi,z'_\phi)$, which specify the tangent plane at a point $(\theta,\phi)$, normalizes them to have the unit length, and then computes their vector product.}
\bea\label{frame}
&&
{\vec e}_\theta =
\begin{pmatrix} \cos\theta\cos\phi \\ \cos\theta\sin\phi \\ -\sin\theta \end{pmatrix} \ ,\qquad
{\vec e}_\phi  =
\begin{pmatrix} -\sin\phi \\ \cos\phi \\ 0 \end{pmatrix} \ , \qquad {\vec e}_r=
\begin{pmatrix} \sin\theta\cos\phi \\ \sin\theta\sin\phi \\ \cos\theta \end{pmatrix} \ ,
\eea
such that $\vec{L}\= p_\theta \vec{e}_\phi - \frac{p_\phi}{\sin\theta} \vec{e}_\theta$, introduces the spin vector \cite{GL}
\be\label{sv}
\vec{J}=g_1 J_1 {\vec e}_r+g_2 J_2 {\vec e}_\theta+g_3 J_3 {\vec e}_\phi, \qquad {\vec{J}}^2=g_1^2 J_1^2+g_2^2 J_2^2+g_3^2 J_3^2,
\ee
where $(g_1,g_2,g_3)$ are nonzero constants (moments of inertia), and finally extends the orbital angular momentum in (\ref{H}) to include the spin part 
\be\label{fv}
\vec{\cJ}=\vec{L}+\vec{J}.
\ee
After that, one demands $\cJ_i$ to obey the $su(2)$ structure relations and identifies the Hamiltonian of a spinning particle on $\mathcal{S}^2$ with the Casimir element
\be\label{Hami}
H=\frac 12  \cJ_i \cJ_i=\frac 12 \left(p_\theta^2+\frac{p_\phi^2}{\sin^2{\theta}} \right)+\frac 12 \left(g_1^2 J_1^2+g_2^2 J_2^2+g_3^2 J_3^2 \right)-\frac{g_2 J_2 p_\phi}{\sin{\theta}}+g_3 J_3 p_\theta.
\ee
Note that discarding the spin degrees of freedom one reproduces the Hamiltonian of a free particle on $\mathcal{S}^2$. Omitting the angular variables, one gets the Hamiltonian typical for describing $3d$ rigid body dynamics \cite{AP}. 

From the equations $\{\cJ_i,\cJ_j \}=\epsilon_{ijk} \cJ_k$ one finds three Poisson brackets
\bea\label{br1}
&&
\{p_\theta,p_\phi\}=-g_1 J_1 \sin{\theta}-g_2 J_2 \cos{\theta}-g_2 g_3 \{J_2,J_3 \} \sin{\theta}+g_2 \{p_\theta,J_2\} \sin{\theta}+g_3 \{p_\phi,J_3 \},
\nonumber\\[6pt]
&&
\{p_\theta,J_1\}=g_3 \{J_1,J_3\}, \qquad  \{p_\phi,J_1 \}=-g_2 \{J_1,J_2\}\sin{\theta},
\eea
with $\{p_\theta,J_2\}$ and $\{p_\phi,J_3 \}$ to be fixed below. In obtaining Eqs. (\ref{br1}), we used the fact that the triple $({\vec e}_r,{\vec e}_\theta,{\vec e}_\phi)$ is closed under the Poisson action of $(p_\theta,p_\phi)$
\begin{align}
&
 \{p_\theta, {\vec e}_\theta \}={\vec e}_r, && \{p_\theta, {\vec e}_\phi \}=0, && \{p_\theta, {\vec e}_r \}=-{\vec e}_\theta,
\nonumber\\[6pt]
&
\{p_\phi, {\vec e}_\theta \}=-\cos{\theta}{\vec e}_\phi, &&\{p_\phi, {\vec e}_\phi \}=\sin{\theta}  {\vec e}_r+\cos{\theta}  {\vec e}_\theta, && \{p_\phi, {\vec e}_r \}=-\sin{\theta} {\vec e}_\phi,
\end{align}
and took into account the identities
\bea
&&
{\left(e_r \right)}_i {\left(e_\theta \right)}_j-{\left(e_r \right)}_j {\left(e_\theta \right)}_i=\epsilon_{ijk} {\left(e_\phi \right)}_k, \qquad
{\left(e_r \right)}_i {\left(e_\phi \right)}_j-{\left(e_r \right)}_j {\left(e_\phi \right)}_i=-\epsilon_{ijk} {\left(e_\theta \right)}_k,
\nonumber\\[2pt]
&&
{\left(e_\theta \right)}_i {\left(e_\phi \right)}_j-{\left(e_\theta \right)}_j {\left(e_\phi \right)}_i=\epsilon_{ijk} {\left(e_r \right)}_k.
\eea

It is straightforward to verify that the second and third brackets in (\ref{br1}) ensure the relations
\be
\{J_1, \cJ_i \}=0.
\ee
Thus, at this stage the system is characterised by three functionally independent integrals of motion in involution $H$, $\cJ_3$ and $J_1$.

It seems reasonable to fix the remaining Poisson brackets from the requirement that the ensuing spinning particle on $\mathcal{S}^2$ be integrable.
Focusing on the spin sector, one has two equations of motion in an unparametrized form and hence one more function commuting with $(H,\cJ_3,J_1)$ is needed in order to provide the Liouville integrability. From the group--theoretic standpoint, a natural choice is the Casimir element $\mathcal{I}$ exposed above in Table 1. Demanding $\{\mathcal{I},\cJ_3 \}=0$ and $\{\mathcal{I},H\}=0$, one obtains
\bea\label{br2}
&&
\{p_\phi,\mathcal{I} \}=\{p_\phi,J_i \} \frac{\partial \mathcal{I}}{\partial J_i}=0, \qquad \{p_\theta,\mathcal{I} \}=\{p_\theta,J_i \} \frac{\partial \mathcal{I}}{\partial J_i}=0.
\eea
Depending on a $3d$ real Lie algebra at hand, these equations allow one to express two brackets in terms of the other. Taking into account (\ref{br1}), one concludes that two brackets among $(p_\theta,p_\phi)$ and $J_i$ are still missing.

As the final step, one requires the Jacobi identities to hold, which ultimately give\footnote{The momenta $(p_\theta,p_\phi)$ entering the $su(2)$ generators in (\ref{H}) are conventionally defined up to a pure gauge vector potential $p_\theta \to p_\theta+A_\theta(\theta,\phi)$, $p_\phi \to p_\phi+A_\phi(\theta,\phi)$, $\partial_\theta A_\phi-\partial_\phi A_\theta=0$. When analysing the Jacobi identities, we discarded a pure gauge vector field contributions. }
\bea\label{br3}
&&
\{p_\theta,p_\phi\}=-g_1 J_1 \sin{\theta}-g_2 J_2 \cos{\theta}-g_1 g_3 \{J_1,J_3 \} \cos{\theta}+g_2 g_3 \{J_2,J_3 \} \sin{\theta},
\nonumber\\[6pt]
&&
\{p_\theta,J_1\}=g_3 \{J_1,J_3\}, \qquad \quad \quad
\{p_\theta,J_2\}=g_3 \{J_2,J_3\}, \qquad \quad \quad \{p_\theta,J_3\}=0,
\nonumber\\[2pt]
&&
\{p_\phi,J_1 \}=-g_2 \{J_1,J_2\}\sin{\theta}, \quad \{p_\phi,J_2 \}=-g_1 \{J_1,J_2\}\cos{\theta},
\nonumber\\[6pt]
&&
\{p_\phi,J_3 \}=-g_1 \{J_1,J_3\}\cos{\theta}+g_2 \{J_2,J_3\}\sin{\theta}.
\eea
This completes our construction of a Hamiltonian formulation for a spinning particle on $\mathcal{S}^2$, the spin degrees of freedom of which are described by a $3$--vector obeying the structure relations of a generic $3d$ real Lie algebra. For the reader's convenience, we expose the Poisson brackets (\ref{br3}) for each instance in the Bianchi classification as well as the second invariant $\mathcal{I}$ in Appendix.

Note that the resulting system is minimally superintegrable as the Liouville integrals of motion $(H,\cJ_3,J_1,\mathcal{I})$ can be extended to include $\cJ_2$ (or alternatively $\cJ_1$). It is straightforward to verify that $(H,\cJ_3,J_1,\mathcal{I},\cJ_2)$ are functionally independent, the only nonzero bracket being $\{\cJ_2,\cJ_3\}$.

It might seem odd that the Poisson brackets among the spin degrees of freedom $J_i$ and the angular variables $(p_\theta,p_\phi)$ are not canonical. Yet, it is worth recalling
the Poisson structure underlying a general relativistic spinning particle on a curved background \cite{AKH}
\be \label{Br}
\bal
&
\{x^\mu,p_\nu \}={\delta^\mu}_\nu, \quad \{p_\mu,p_\nu\}=-\sfrac 12 R_{\mu\nu\lambda\sigma} S^{\lambda\sigma}, \quad \{S^{\mu\nu},p_\lambda \}=\Gamma^\mu_{\lambda\sigma} S^{\nu\sigma}-\Gamma^\nu_{\lambda\sigma} S^{\mu\sigma},
\\[2pt]
&
\{S^{\mu\nu},S^{\lambda\sigma} \}=g^{\mu\lambda} S^{\nu\sigma}+g^{\nu\sigma} S^{\mu\lambda}-g^{\mu\sigma} S^{\nu\lambda}-g^{\nu\lambda} S^{\mu\sigma},
\eal
\ee
where $(x^\mu,p_\mu)$, $\mu=0,1,2,3$, are the canonical variables, $S^{\mu\nu}=-S^{\nu\mu}$ are the spin degrees of freedom,  $g^{\mu\nu}$ is the inverse metric tensor, $\Gamma^\mu_{\lambda\sigma}$ are the Christoffel symbols, and $R_{\mu\nu\lambda\sigma}$ is the Riemann tensor. Focusing on a spherically symmetric background and properly reducing the corresponding $su(2)$ Killing vector fields, one arrives at a Bianchi type--IX spinning particle on $\mathcal{S}^2$ \cite{GL}, which is a particular member in the set of models constructed above.

\vspace{0.5cm}

\noindent
{\bf 4. Qualitative dynamics}\\

\noindent
Let us briefly discuss a qualitative dynamical behaviour of the integrable systems built in the preceding section.
Passing to the Cartesian coordinates $\vec x=(\sin\theta\cos\phi,\sin\theta\sin\phi,\cos\theta)$, one gets the relations
\be\label{cone}
(\vec x, \vec\cJ)=g_1 J_1, \qquad \dot{x}_i\dot{x}_i \= {\dot\theta}^2+{\dot\phi}^2 \sin^2{\theta}=2H-g_1^2 J_1^2,
\ee
which imply a uniform motion along a circular orbit on $\mathcal{S}^2$, which is an intersection of the cone and the sphere.
The apex semi--angle of the cone depends on the energy of the full system and the conserved spin component $J_1$
\be\label{angle}
\cos{\alpha}=\frac{g_1 J_1}{\sqrt{2H}}.
\ee
If $g_1 J_1=0$ the cone opens to the plane $x_i \mathcal{J}_i=0$, and the orbit becomes a great circle.

Turning to the spin sector, one reveals a conserved component $J_1=\mbox{const}$, while $J_2$ and $J_3$ swing in a way dependent on the angular variables.
The analysis becomes more transparent if one changes $(p_\theta,p_\phi)$ so as to partially diagonalize the brackets
\bea
&&
p'_\theta=p_\theta+g_3 J_3, \qquad p'_\phi=p_\phi+g_1 J_1 \cos{\theta}-g_2 J_2 \sin{\theta},
\nonumber\\[2pt]
&&
\{p'_\theta,J_i \}=0, \qquad \quad \{p'_\phi,J_i \}=0, \qquad \quad \{p'_\theta,p'_\phi \}=0.
\eea
Then (\ref{Hami}) simplifies to
\be\label{Hnew}
H'=\frac 12 \left({p'_\theta}^2+\frac{{\left(p'_\phi-g_1 J_1 \cos{\theta} \right)}^2}{\sin^2{\theta}}  +g_1^2 J_1^2\right).
\ee
Remarkably enough, $H'$ coincides with the Hamiltonian of a particle on $\mathcal{S}^2$ in the presence of a magnetic monopole field, the magnetic charge being promoted to the conserved spin component $J_1$ multiplied by a constant $g_1$. In this coordinate system it becomes evident that the evolution of $(\theta,\phi)$ is identical to that of a particle on $\mathcal{S}^2$ coupled to an external field of the Dirac monopole, while $J_2$, $J_3$ satisfy the equation
\be\label{es}
{\dot J}_{2,3}=\frac{g_1}{\sin^2{\theta}} \left(p'_\phi \cos{\theta}-g_1 J_1\right) \{J_1,J_{2,3} \}.
\ee

Having solved the equations of motion for the angular variables, one can redefine the temporal parameter
\be
\tau(t)=g_1 \int dt \left( \frac{p'_\phi (t) \cos{\theta}(t)-g_1 J_1 }{\sin^2{\theta(t)}}\right)   +\tau_0,
\ee
where $\tau_0$ is a constant, and reduce (\ref{es}) to the linear equations
\be
J'_{2,3}=\{J_1,J_{2,3} \},
\ee
the prime indicating the derivative with respect to $\tau$, which can be easily integrated. The results are given below in Table 2, in which $C_1$ and $C_2$ denote constants of integration. Because the motion on $\mathcal{S}^2$ is periodic, so is the swinging of the vector with components $(J_2(t),J_3(t))$ in the tangent plane.
\newpage
\begin{center}
Table 2. Evolution of the spin degrees of freedom with time
\end{center}
\begin{eqnarray*}
\footnotesize
\begin{array}{|l|r|r|r|r|r|r|r|r|c|}
\hline
  & J_2(t)  & J_3(t)   \\
  \hline
~ \mbox{type II}  & C_1 & C_2  \\
\hline
\mbox{ type III}  & C_1 \left(1 + e^{2 \tau(t)}\right) + C_2 \left(1 - e^{2 \tau(t)}\right)  & C_1 \left(1 - e^{2 \tau(t)}\right)  + C_2 \left(1 + e^{2 \tau(t)}\right) \\
\hline
\mbox{ type IV} & C_1 e^{\tau(t)} + C_2 \tau(t) e^{\tau(t)} & C_2 e^{\tau(t)}\\
\hline
\mbox{ type V} & C_1 e^{\tau(t)} & C_2 e^{\tau(t)} \\
\hline
\mbox{ type VI} & C_1 \left(e^{(a-1)\tau(t)} + e^{(a+1) \tau(t)}\right)   & C_1 \left(e^{(a-1)\tau(t)} - e^{(a+1) \tau(t)}\right)  \\
& + C_2 \left(e^{(a-1)\tau(t)}- e^{(a+1) \tau(t)}\right) & + C_2 \left(e^{(a-1)\tau(t)}+e^{(a+1) \tau(t)}\right) \\
\hline
~ \mbox{type $$VI$_0$} & C_1 & C_1 \tau(t)+C_2  \\
\hline
 \mbox{ type VII} & \left(C_1 \cos{\tau(t)}+C_2 \sin{\tau(t)} \right) e^{a \tau(t)} & -\left(C_1 \sin{\tau(t)}-C_2 \cos{\tau(t)} \right) e^{a \tau(t)} \\
\hline
\mbox{ type $$VII$_0$} & C_1 & -C_1 \tau(t)+C_2 \\
\hline
\mbox{ type VIII} & C_1 \cosh{\tau(t)}+C_2 \sinh{\tau(t)} & -C_1 \sinh{\tau(t)}-C_2 \cosh{\tau(t)} \\
\hline
 \mbox{ type IX} & C_1 \cos{\tau(t)}+C_2 \sin{\tau(t)} & -C_1 \sin{\tau(t)}+C_2 \cos{\tau(t)} \\
\hline
\end{array}
\end{eqnarray*}
\vspace{0.2cm}

Concluding this section, we note that the Hamiltonian (\ref{Hnew}) offers another perspective on the material in the preceding section. Consider a particle on $\mathcal{S}^2$ coupled to an external field of the Dirac monopole. The system is known to be $su(2)$ invariant and the corresponding generators involve the magnetic charge $q$
\be\label{MM}
\vec{L}'=
\begin{pmatrix} -p_\theta \sin{\phi}-p_\phi \cot{\theta} \cos{\phi}+q \cos{\phi} \sin^{-1}{\theta}, \\ p_\theta \cos{\phi}-p_\phi \cot{\theta} \sin{\phi}+q \sin{\phi} \sin^{-1}{\theta} \\ p_\phi \end{pmatrix}, \qquad \{L'_i,L'_j \}=\epsilon_{ijk} L'_k.
\ee
Let us choose a $3d$ real Lie algebra and use its structure constants to specify the (degenerate) Poisson bracket $\{J_i,J_j \}=c_{ij}^k J_k$. Implementing the oxidation with respect to $q$
\be\label{ox}
q  ~ \rightarrow ~ g_1 J_1,
\ee
where $g_1$ is a constant, and taking the Casimir element of $su(2)$ to be the Hamiltonian of the extended system $H=\frac 12 \vec{L}'^2$, one obtains a spinning extension for which $J_1$ is a linear integral of motion. 

In Ref. \cite{PSWZ}, the Casimir invariants were found for all real algebras of dimension up to five and for all nilpotent real algebras of dimension six. Implementing the oxidation (\ref{ox}), one automatically gets similar integrable extensions of a free particle on $\mathcal{S}^2$, the internal degrees of freedom of which satisfy a $d=4,5,6$ real Lie algebra. $D(2,1;a)$ supersymmetrization then follows as described in Sec. 2.

\vspace{0.5cm}

\noindent
{\bf 5. Generalisations}\\

\noindent
Let us discuss some directions in which the analysis in Sec. 3 can be generalised.

When decomposing the spin vector $\vec{J}$ on the basis $({\vec e}_r,{\vec e}_\theta,{\vec e}_\phi)$ in Eq. (\ref{sv}) above, we chose $J_1$ to be a companion of ${\vec e}_r$. It was later established that $J_1$ is a constant of the motion of the spinning particle on $\mathcal{S}^2$, while the interchange of $J_2$ and $J_3$ affects the resulting system only slightly. In a similar fashion one could build models in which either $J_2$ or $J_3$ would be a linear integral of motion. As follows from Table 1, in the latter two cases the Casimir element would degenerate either to linear or quadratic integral of motion depending on the item in the Bianchi classification.

The models in Sec. 3 can be readily coupled to an external field of the Dirac monopole. It suffices to extend (\ref{fv}) 
\be\label{mm}
\vec\cJ \quad \rightarrow \quad \vec\cJ'=\vec\cJ+q \vec B, \qquad \vec B=\begin{pmatrix} \cos{\phi} \sin^{-1}{\theta} \\  \sin{\phi} \sin^{-1}{\theta} \\ 0 \end{pmatrix},
\ee
where $q$ is the magnetic charge.
The presence of the external field alters the dynamics only slightly. For the orbital motion one finds
\be
(\vec x, \vec\cJ')=g_1 J_1+q, \qquad {\dot\theta}^2+{\dot\phi}^2 \sin^2{\theta}= 2H-{(g_1 J_1+q)}^2.
\ee
In particular, at $g_1 J_1+q=0$ the impact of the spin degrees of freedom on the orbital motion on $\mathcal{S}^2$ is compensated by the external field.

The model (\ref{mm}) can be further generalised by introducing into the consideration an extra canonical pair $(\chi,p_\chi)$, obeying the standard Poisson bracket $\{\chi,p_\chi \}=1$, and implementing the oxidation with respect to $q$
\be
q  ~ \rightarrow ~ p_\chi.
\ee
The corresponding $su(2)$ generators are the building blocks to construct a spinning particle propagating on the group manifold of $SU(2)$. Its Liouville integrability is provided by five
functionally independent integrals of motion in involution $(H,\mathcal{J}_3,J_1,\mathcal{I},p_\chi)$. 

As was mentioned after Eq. (\ref{angle}), for $g_1=0$ the orbit on $\mathcal{S}^2$ is a great circle. In this case one can add external scalar potential without spoiling the minimal superintegrability. It suffices to consider three functions
\bea\label{II}
&&
I_1=\mathcal{J}_1^2+{\left(\nu_1 \sin^{-1}{\phi}\cot{\theta}+\nu_3 \sin{\phi} \tan{\theta}\right)}^2,
\nonumber\\[2pt]
&&
I_2=\mathcal{J}_2^2+{\left(\nu_2 \cos^{-1}{\phi}\cot{\theta}+\nu_3 \cos{\phi} \tan{\theta}\right)}^2,
\nonumber\\[2pt]
&&
I_3=\mathcal{J}_3^2+{\left(\nu_1 \cot{\phi}+\nu_2 \tan{\phi} \right)}^2,
\eea
where $(\nu_1,\nu_2,\nu_3)$ are (coupling) constants and verify that they commute with the Hamiltonian
$H=\frac 12 \left(I_1+I_2+I_3\right)$.
Four functionally independent integrals of motion in involution include $(H,I_1,\mathcal{I},J_1)$. Adding $I_2$ (or alternatively $I_3$) renders the model minimally superintegrable.

\vspace{0.5cm}

\noindent
{\bf 6. Conclusion}\\

\noindent
To summarise, in this work we have constructed a minimally superintegrable spinning particle on $\mathcal{S}^2$, the spin degrees of freedom of which are represented by a $3$--vector obeying the structure relations of a $3d$ real Lie algebra in accord with the Bianchi classification. Generalisations involving an external field of the Dirac monopole, or the motion on the group manifold of $SU(2)$, or a scalar potential giving rise to two quadratic constants of the motion were proposed. It was argued that similar integrable extensions, the internal degrees of freedom of which satisfy $d=4,5,6$ real Lie algebra, can be constructed by considering a particle on $\mathcal{S}^2$ coupled to an external field of the Dirac monopole and implementing an oxidation with respect to the magnetic charge.

As a possible further development, it would be interesting to study whether integrable spinning extensions of a particle on $\mathcal{S}^2$ (or $\mathcal{S}^3$) can be constructed beyond the oxidation scheme in Sec. 4, i.e. avoiding a linear integral of motion in the spin sector.

\vspace{0.5cm}

\noindent{\bf Acknowledgements}\\

\noindent
This work is supported by the Russian Science Foundation, grant No 19-11-00005.

\vspace{0.5cm}

\noindent
{\bf Appendix}

\vspace{0.5cm}
\noindent
In this Appendix, we display the Poisson brackets among $p_\theta$, $p_\phi$, and $J_i$ for each instance in the Bianchi classification. We omit the abelian type--I case as it is of little physical interest. For the type--II case, $J_1$ coincides with the Casimir invariant $\mathcal{I}$. In order to provide integrability, $J_2$ was chosen to be the second integral of motion.
\vspace{0.2cm}
\begin{eqnarray*}
\footnotesize
\begin{array}{|l|r|r|r|r|r|r|r|r|c|}
\hline
                    & \mbox{type II} &  \mbox{ type III} & \mbox{ type IV} &  \mbox{ type V} &  \mbox{ type VI} \\
\hline
\{p_\theta,J_1\}    & 0 & -g_3(J_2-J_3) &   g_3 J_3 &  g_3 J_3 &  -g_3 (J_2-a J_3) \\
\hline
\{p_\theta,J_2\}    & g_3 J_1 & 0 &  0 &  0 &  0 \\
\hline
\{p_\theta,J_3\}    & 0 & 0 &  0 &  0 &  0\\
\hline
\{p_\phi,J_1\}      & 0 & -g_2(J_2-J_3)\sin{\theta} & -g_2(J_2+ J_3) \sin{\theta} &  -g_2 J_2 \sin{\theta} & -g_2 (a J_2-J_3) \\
                    & & & & & \times \sin{\theta} \\
\hline
\{p_\phi,J_2\}      & 0 & -g_1(J_2-J_3)\cos{\theta} &  -g_1 (J_2+J_3) \cos{\theta} &  -g_1 J_2 \cos{\theta} &  -g_1 (a J_2-J_3) \\
                    & & & & & \times \cos{\theta}\\
\hline
\{p_\phi,J_3\}      & g_2 J_1 \sin{\theta} & g_1(J_2-J_3)\cos{\theta} &  -g_1 J_3 \cos{\theta} & -g_1 J_3 \cos{\theta} &  g_1 (J_2-a J_3) \\
                    & & & & & \times \cos{\theta}\\
\hline
\{p_\theta,p_\phi\} & (g_2 g_3-g_1)  & -g_1 J_1 \sin{\theta}  &  -g_1 J_1 \sin{\theta} & -g_1 J_1 \sin{\theta} &  -g_1 J_1 \sin{\theta} \\
%\hline
                    & \times J_1 \sin{\theta}& +(g_1 g_3-g_2) J_2 \cos{\theta}  & -g_2 J_2 \cos{\theta} &  -g_2 J_2 \cos{\theta} &  +(g_1 g_3-g_2)  \\
                    &  -g_2 J_2 \cos{\theta}  & -g_1 g_3 J_3 \cos{\theta}  & -g_1 g_3 J_3 \cos{\theta}& -g_1 g_3 J_3 \cos{\theta} &  \times J_2 \cos{\theta}          \\
                    & & & & & -a g_1 g_3 J_3 \cos{\theta} \\
\hline
~  \mathcal{I}     &   J_2    &   J_2+J_3 &         \frac{J_2}{J_3}-\ln{J_3}           &       \frac{J_2}{J_3}                 & J_3^2 {\left(1+\frac{J_2}{J_3} \right)}^{1+a}    \\
                     & & & & & \times {\left(1-\frac{J_2}{J_3} \right)}^{1-a} \\
\hline
\end{array}
\end{eqnarray*}
\begin{eqnarray*}
\footnotesize
\begin{array}{|l|r|r|r|r|r|r|r|r|c|}
\hline
                    &\mbox{type $$VI$_0$} &  \mbox{ type VII} & \mbox{ type $$VII$_0$} &  \mbox{ type VIII} &  \mbox{ type IX} \\
\hline
\{p_\theta,J_1\}    & g_3 J_2 & -g_3 (J_2-a J_3) &   -g_3 J_2 & -g_3 J_2 &  -g_3 J_2 \\
\hline
\{p_\theta,J_2\}    & g_3 J_1 & 0 &  g_3 J_1 &  g_3 J_1 &  g_3 J_1 \\
\hline
\{p_\theta,J_3\}    & 0 & 0 &  0&  0 &  0\\
\hline
\{p_\phi,J_1\}      & 0 & -g_2(a J_2+J_3) \sin{\theta} & 0 &  g_2 J_3 \sin{\theta} & -g_2 J_3 \sin{\theta} \\
\hline
\{p_\phi,J_2\}      & 0 & -g_1(a J_2+J_3) \cos{\theta} &  0 &  g_1 J_3 \cos{\theta}&  -g_1 J_3 \cos{\theta} \\
\hline
\{p_\phi,J_3\}      & g_2 J_1 \sin{\theta} & g_1(J_2- a J_3)\cos{\theta} &  g_2 J_1 \sin{\theta} & g_2 J_1 \sin{\theta} &  g_2 J_1 \sin{\theta} \\
                    & -g_1 J_2 \cos{\theta} &                            & +g_1 J_2 \cos{\theta}                      &   +g_1 J_2 \cos{\theta}& +g_1 J_2 \cos{\theta} \\
\hline
\{p_\theta,p_\phi\} & (g_2 g_3-g_1)  & -g_1 J_1 \sin{\theta} &  (g_2 g_3-g_1) J_1 \sin{\theta}  & (g_2 g_3-g_1) J_1 \sin{\theta} &  (g_2 g_3-g_1)  \\
                    &   \times J_1 \sin{\theta} &  +(g_1 g_3-g_2) J_2 \cos{\theta} & +(g_1 g_3-g_2) J_2 \cos{\theta} &  +(g_1 g_3-g_2) J_2\cos{\theta} & \times J_1 \sin{\theta}   \\
                    & -(g_1 g_3+g_2)  & -a g_1 g_3 J_3 \cos{\theta} & & & +(g_1 g_3-g_2) \\
                     &\times J_2 \cos{\theta}   & & & & \times  J_2 \cos{\theta} \\
                    \hline
~  \mathcal{I}       &   J_2   &   (J_2^2 + J_3^2)  &         J_2              &       J_2^2-J_3^2                 & J_2^2+J_3^2     \\
                     &  & \times e^{-2 a \arctan{\frac{J2}{J3}}}  & & & \\
\hline
\end{array}
\end{eqnarray*}


\begin{thebibliography}{nn}
\bibitem{FIL}
S. Fedoruk, E. Ivanov, O. Lechtenfeld, {\it Superconformal mechanics}, J. Phys. A {\bf 45} (2012) 173001, arXiv:1112.1947.
\bibitem{FIL1}
S. Fedoruk, E. Ivanov, O. Lechtenfeld, {\it Supersymmetric Calogero models by gauging}, Phys. Rev. D {\bf 79} (2009) 105015, arXiv:0812.4276.
\bibitem{FIL2}
S. Fedoruk, E. Ivanov, O. Lechtenfeld, {\it New $D(2,1;\alpha)$ mechanics with spin variables}, JHEP {\bf 1004} (2010) 129, arXiv:0912.3508.
\bibitem{KL}
S. Krivonos, O. Lechtenfeld, {\it Many-particle mechanics with $D(2,1;\alpha)$ superconformal symmetry}, JHEP {\bf 1102} (2011) 042, arXiv:1012.4639.
\bibitem{KLS}
S. Krivonos, O. Lechtenfeld, A. Sutulin, {\it $\mathcal{N}$--extended supersymmetric Calogero models}, Phys. Lett. B {\bf 784} (2018) 137, arXiv:1804.10825.
\bibitem{KLS1}
S. Krivonos, O. Lechtenfeld, A. Sutulin, {\it Supersymmetric many--body Euler-–Calogero-–Moser model}, Phys. Lett. B {\bf 790} (2019) 191, arXiv:1812.03530.
\bibitem{GL}
A. Galajinsky, O. Lechtenfeld,  {\it Spinning extensions of $D(2,1;\alpha)$ superconformal mechanics}, JHEP {\bf 1903} (2019) 069, arXiv:1902.06851.
\bibitem{FIL3}
S. Fedoruk, E. Ivanov, O. Lechtenfeld, {\it Supersymmetric hyperbolic Calogero--Sutherland models by gauging}, Nucl. Phys. B {\bf 944} (2019) 114633, arXiv:1902.08023.
\bibitem{F}
S. Fedoruk, {\it $\mathcal{N}=2$ supersymmetric hyperbolic Calogero--Sutherland model}, Nucl. Phys. B {\bf 953} (2020) 114977, arXiv:1910.07348.
\bibitem{G}
A. Galajinsky, {\it  Bianchi type--$V$ spinning particle on $\mathcal{S}^2$}, JHEP {\bf 2003} (2020) 143, arXiv:1912.13339.
\bibitem{IPW}
L. Inzunza, M.S. Plyushchay, A. Wipf, {\it Hidden symmetry and (super)conformal mechanics in a monopole background}, JHEP {\bf 2004} (2020) 028, arXiv:2002.04341.
\bibitem{F1}
S. Fedoruk, {\it $\mathcal{N}=4$ supersymmetric $U(2)$--spin hyperbolic Calogero--Sutherland model}, Nucl. Phys. B {\bf 961} (2020) 115234, arXiv:2007.11424.
\bibitem{B}
L. Bianchi, {\it Sugli spazi a tre dimensioni che ammettono un gruppo continuo di movimenti},
Memorie di Matematica e di Fisica della Societa Italiana delle Scienze, Serie Terza, {\bf 11} (1898) 267.
\bibitem{G1}	
A. Galajinsky, {\it Couplings in $D(2,1;\alpha)$ superconformal mechanics from the $SU(2)$ perspective},
JHEP {\bf 1703} (2017) 054, arXiv:1702.01955.
\bibitem{DNF}
B.A. Dubrovin, A.T. Fomenko, S.P. Novikov, {\it Modern geometry – methods and applications. Part I. The geometry of surfaces, transformation groups, and fields}. Graduate Texts in Mathematics, Vol. 93, Springer-Verlag, New York, 1984.
\bibitem{AP}
A.M. Perelomov, {\it Integrable systems of classical mechanics and Lie algebras}, Birkh\"auser Basel, 1990.
\bibitem{AKH}
G. d'Ambrosi, S. Satish Kumar, J.W. van Holten, {\it Covariant hamiltonian spin dynamics in curved space-time},
Phys. Lett. B {\bf 743} (2015) 478, arXiv:1501.04879.
\bibitem{PSWZ}
J. Patera, R.T. Sharp, P. Winternitz, H. Zassenhaus, {\it Invariants of real low dimension Lie algebras}, J. Math. Phys. {\bf 17} (1976) 986.
\end{thebibliography}
\end{document}